\documentclass{PoS}

\usepackage{amsmath}

\def\dzdzb{{~\raise.85em\hbox{{\tiny{(}\textemdash\tiny{)}}}\kern-1.05em\lower0.0em\hbox{$D^0$}~}}
\def\bsbsb{{~\raise.85em\hbox{{\tiny{(}\textemdash\tiny{)}}}\kern-1.05em\lower0.0em\hbox{$B_s^0$}~}\xspace}

\def\bstodskpipi{\overline{B^0_s}\to D_s^{\pm}K^{\mp}\pi^{\pm}\pi^{\mp}}
\def\btodpipipi{\overline{B^0}\to D^+\pi^-\pi^+\pi^-}
\def\btodkpipi{\overline{B^0}\to D^+K^-\pi^+\pi^-}
\def\btodzeropipipi{B^-\to D^0\pi^-\pi^+\pi^-}
\def\btodzerokpipi{B^-\to D^0 K^-\pi^+\pi^-}

\def\btodh{B\to Dh}

\def\bstodsk{\overline{B^0_s}\to D_s^{\pm}K^{\mp}}
\def\btodpi{\overline{B}^0\to D^+\pi^-}
\def\btodk{\overline{B^0}\to D^+K^-}

\def\btodzerok{B^-\to D^0 K^-}

\def\eff{\epsilon}
\def\btodzerokstar{\overline{B^{0}}\to D^0 \overline{K^{*0}}}
\def\bstodzerokstar{{\overline{B^0_s}}\to D^0 K^{*0}}
\def\btodzerorho{{\overline{B^{0}}}\to D^0\rho^{0}}

\def\ifb{\rm fb^{-1}}
\def\ipb{\rm pb^{-1}}
\def\br{{\mathcal{B}}}

\vspace*{-1.5cm}

\title{$B_{(s)}\to D_{(s)}h(h)(h)$ Decays in LHCb}

\ShortTitle{$B_{(s)}\to D_{(s)}h(h)(h)$ Decays in LHCb}

\author{\speaker{Steven R. Blusk}\thanks{On behalf of the LHCb Collaboration}\\
        Syracuse University\\
        E-mail: \email{sblusk@syr.edu}}

\abstract{We report recent measurements from LHCb on $B_{(s)}\to D_{(s)}h(h)(h)$ decays 
using $\sim$35~$\ipb$ of data collected in 2010. In brief, we measure the following
ratios of branching fractions: 

\begin{align*}
\dfrac{\br(\btodk)}{\br(\btodpi)} &= 0.0752\pm0.0064\pm0.0026 \\
\dfrac{\br(\bstodzerokstar)}{\br(\btodzerorho)} &= 1.39\pm0.31\pm0.17\pm0.18 \\
\dfrac{\br(\btodkpipi)}{\br(\btodpipipi)} &= 0.052\pm0.009\pm0.005 \\
\dfrac{\br(\btodzerokpipi)}{\br(\btodzeropipipi)} &= 0.096\pm0.015\pm0.008
\end{align*}

\noindent where the uncertainties are statistical and systematic, respectively. The first
of these measurements is the most precise to date, and the others are first observations. }

\FullConference{The 13th International Conference on B-Physics at Hadron Machines - Beauty2011,\\
		April 04-08, 2011\\
		Amsterdam, The Netherlands}

\begin{document}

\section{Introduction}
One of the key objectives of particle physics is to search for new physics (NP) in the
decays of beauty and charm particles. In the presence of NP, decays containing quantum loops 
would acquire an additional NP amplitude, and the interference between it and the standard model (SM)
amplitudes could give rise to sizeable deviations in the observed rates, angular distributions,
or CP asymmetries. In some cases, the SM provides predictions to which
measurements can be compared, {\it e.g.}, $\br(B_s\to\mu^+\mu^-)=(3.2\pm0.2)\times10^{-9}$~\cite{buras_bs2mumu},
zero-crossing point in $B^0\to K^{*0}\ell^+\ell^-$~\cite{beneke_kstarll}. 


At the heart of weak heavy flavor decays is the CKM matrix~\cite{ckm}, 
describing a Unitary rotation between the flavor eigenstates and the mass eigenstates. Its
four parameters, three real angles and one complex phase are not predicted by the SM and must be measured.
The CKM matrix determines not only the relative strengths of various quark transitions, but allows
for matter-antimatter asymmetries, if the complex phase is non-zero. Applying the unitarity constraint to
$b$-$d$ columns produces a triangle in the complex plane (so-called Unitarity triangle (UT)). 
The sides and angles of
this triangle are related to two of the the four CKM parameters, and each can be probed through a variety of 
decay processes. 

\begin{figure}[ht]
\centering
\includegraphics[width=125mm]{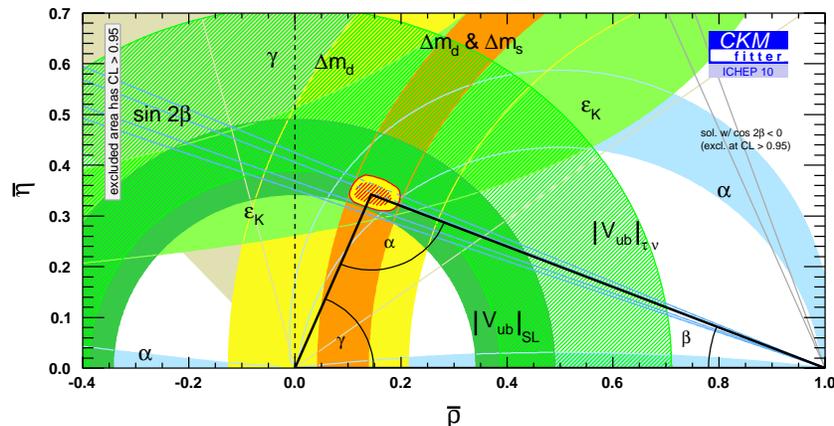}
\vspace{-0.25in}
\caption{Constraints on ($\rho$,$\eta$) from a number of measurements in the flavor sector.}
\label{fig:ckmfit}
\end{figure}

The state of affairs of the Unitarity triangle is shown in Fig.~\ref{fig:ckmfit}. A number of measurements
are combined to determine its apex ($\rho$,$\eta$). 
Since many of these measurements may be influenced
by NP, it is of great importance to precisely determine the apex of this triangle using decays that are both
(expected to be) sensitive and insensitive to NP. Any significant deviation between the apex in the
{\it NP-sensitive}  and {\it NP-insensitive} measurements would be a smoking gun for NP.
There are indications of tension between the various CKM measurements~\cite{soni}.

The least well measured of the angles in the UT is the angle $\gamma$. The current precision
on $\gamma$ ranges from $\sim$11$^o$~\cite{utfit} to 14$^o$~\cite{ckmfitter},
compared to a precision on the other two angles of about 3\% and 4.5\% for $\beta$ and $\alpha$,
respectively. It is therefore a high priority in flavor physics to
make a precise measurement of the angle $\gamma$, and see if the fitted apex in Fig.~\ref{fig:ckmfit}
is consistent with the directly measured value of $\gamma$.

A number of theoretically clean methods for extracting $\gamma$ have been explored in the
literature. Among the most well known techniques are to use the Cabibbo-suppressed (CS) decays 
$\btodzerok$~\cite{ads,glw,ggsz} and $\bstodsk$~\cite{dsk1,dsk2}. 
Beyond these modes, one can exploit higher multiplicity decays, such as
$\btodzerokstar$, $\btodzerokpipi$~\cite{gronau} or $\bstodskpipi$.
Additional sensitivity can be obtained by using $\btodpi$~\cite{b2dpicp} and $\btodpipipi$ decays.
Because these measurements are limited by $|V_{\rm ub}|$, and we only measure $O(10\%)$ of the
charm decays, the rates for these decays are low, requiring a very large data sample.

In 2010, LHCb collected $\sim$35~$\ipb$ of data, about 2\% of a nominal year's luminosity. While this size data
sample is insufficient to begin measurements of $\gamma$, it is sufficient to demonstrate that LHCb can observe
the kinematically similar Cabibbo-favored (CF) decays, with roughly the expected yields (from simulation) and a good
signal-to-background ratio. Here, we report on recent measurements of $B_{(s)}\to D_{(s)}h(h)(h)$ decays ($h=\pi,K$) 
using this data sample.

\section{The LHCb Experiment}
The LHCb experiment is a dedicated flavor experiment at the Large Hadron Collider.
The copious production of $b\bar{b}$ pairs ($\sigma_{b\bar{b}}=284\pm20\pm49~\mu b$~\cite{bxsec}),
combined with the correlated forward production of $b\bar{b}$ pairs, allows
LHCb to trigger on and reconstruct important and rare 
decays with product branching fractions down to $O(10^{-9})$ with a 2~$\ifb$ data sample. 
The detector includes a charged particle tracking system that provides an impact parameter (IP)
resolution of $\sim 16\mu$m + 30$\mu$m/$p_T$ ($p_T$ in GeV/$c$), and a momentum resolution that ranges from 
$\sigma_p/p\sim0.5\%$ at 3 GeV/$c$ to $\sim0.8\%$ at 100 GeV/$c$. Two Ring Imaging Cherenkov Counters (RICH)
provide a kaon particle identification (PID) efficiency of $\sim$95\% for a pion fake rate of a few percent.
Electromagnetic and hadronic calorimeter systems provide for electron and photon identification,
and a muon system provides for muon identification. 
A more detailed description of the LHCb detector can be found in Ref.~\cite{lhcb-det}.

Events are selected by a two level trigger system. The first level, L0, is hardware-based, capable of operating at 40 MHz,
and selects events with either a large transverse energy deposition, $E_T>3.6$~GeV, in the calorimeters, or single/di-muons
detected in the muon system. The output of L0 (up to 1 MHz) is then processed by a High Level Trigger (HLT), which runs 
simplified version of the offline LHCb software. For the analyses presented here, the first level of
the HLT (HLT1) requires at
least one charged particle with $p_T>1.8$~GeV/$c$ and IP>125~$\mu$m~\cite{hlt1}.
A second stage (HLT2), then searches for 2, 3,4-particle vertices using tracks that have $p>5$~GeV/$c$, 
$p_T>0.5$~GeV/$c$ and IP $\chi^2>16$ to any PV (see Ref.~\cite{topo} for more details).
These HLT1 and HLT2 lines each have an efficiency of $\sim$80-90\% for 
a large range of $B$ decays. 
For both L0 and HLT, we can trace offline-reconstructed signal candidates to trigger objects.
Events can then be classified into those in which the event was {\bf T}riggered {\bf O}n {\bf S}ignal (TOS),
or {\bf T}riggered {\bf I}ndependently of the {\bf S}ignal (TIS).

\section{Measurement of $\br(\btodk)$}

The decay $\btodk$ is kinematically similar to $\bstodsk$, which can be used to measure $\gamma$ in a 
time-dependent analysis. By observing $\btodk$ and measuring its rate, we demonstrate LHCb's capabilities
in purely hadronic channels, and it sets the stage for expectations in $\bstodsk$ with a larger data sample.
The decay is measured relative to the kinematically similar $\btodpi$.

The search for this decay starts with selecting particles that are more likely to come from $b-$hadron decay.
Tracks are required to have $p_T>300$~MeV/$c$ and IP~$\chi^2>9$. Using these tracks, candidate $D^+\to K^-\pi^+\pi^+$ 
decays are formed, where we require $\Delta LL(K-\pi)>0$ and $\Delta LL(K-\pi)<10$ for kaons and pions, respectively,
as determined using information from the RICH. Reconstructed $D$ candidates are required to have $p_T>1.5$~GeV/$c$
and vertex $\chi^2/dof<12$. Tighter selections are imposed on bachelor particle candidates, namely
we require $p_T>500$~MeV/$c$ and $\Delta LL(K-\pi)>5$ ($\Delta LL(K-\pi)<0$) for kaons (pions). 
$\btodk$ ($\btodpi$) candidates are formed by combing a $D$ candidate which has invariant mass in the range
$1829<M_{K\pi\pi}<1893$~MeV/$c^2$ with a bachelor kaon (pion) candidate, and requiring it have vertex $\chi^2/dof<12$
and proper time $\tau>0.2$~ps. Events are required to be either TOS or TIS at L0, and 
TOS at HLT. A final boosted-decision-tree, trained on signal MC for signal and sidebands in data for the
background, is used to improve the signal-to-background ratio.
Reconstructed $B^0$ candidates passing these selections are shown in
Fig.~\ref{fig:b2dh}. Signal yields are extracted from an unbinned maximum likelihood fit that includes 
a double-Crystal Ball signal shape~\cite{cbal2}, and shapes to describe background sources from both known
decays, such as $D^*h$, $D\rho$, and random combinations. The fitted yields are $4109\pm75$ $\btodpi$ and
$253\pm21$ $\btodk$ signal events. The ratio of efficiencies, $\eff_{\btodpi}/\eff_{\btodk}=1.221\pm0.007$,
where the departure from unity is mainly driven by the lower PID efficiency for the bachelor kaon in $\btodk$. 
The resulting ratio of branching fractions is measured to be:

\begin{equation}
\frac{\br(\btodk)}{\br(\btodpi)} = 0.0752\pm0.0064\pm0.0026 \nonumber
\end{equation}

\begin{figure}[ht]
\centering
\includegraphics[width=75mm]{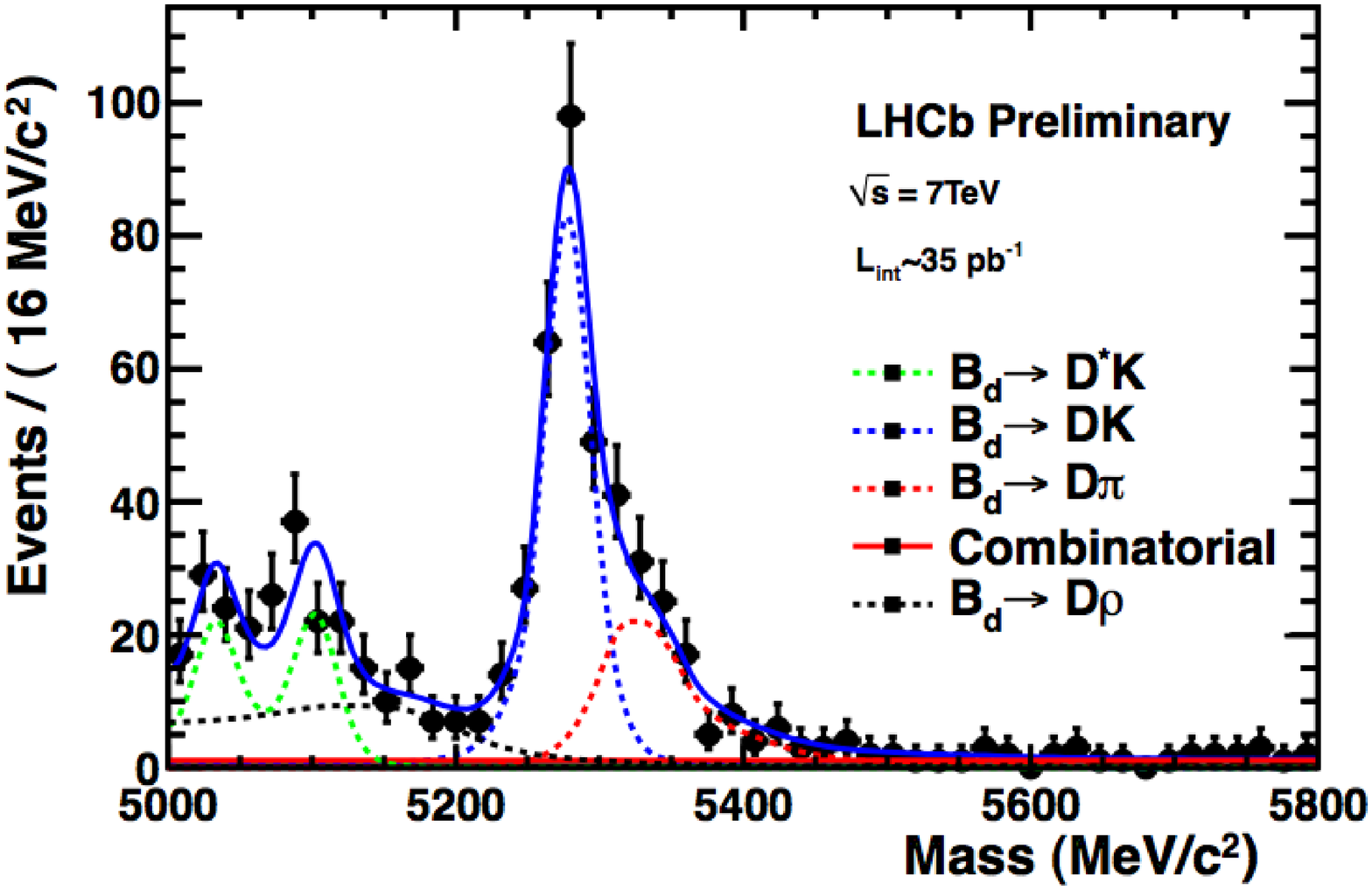}
\includegraphics[width=75mm]{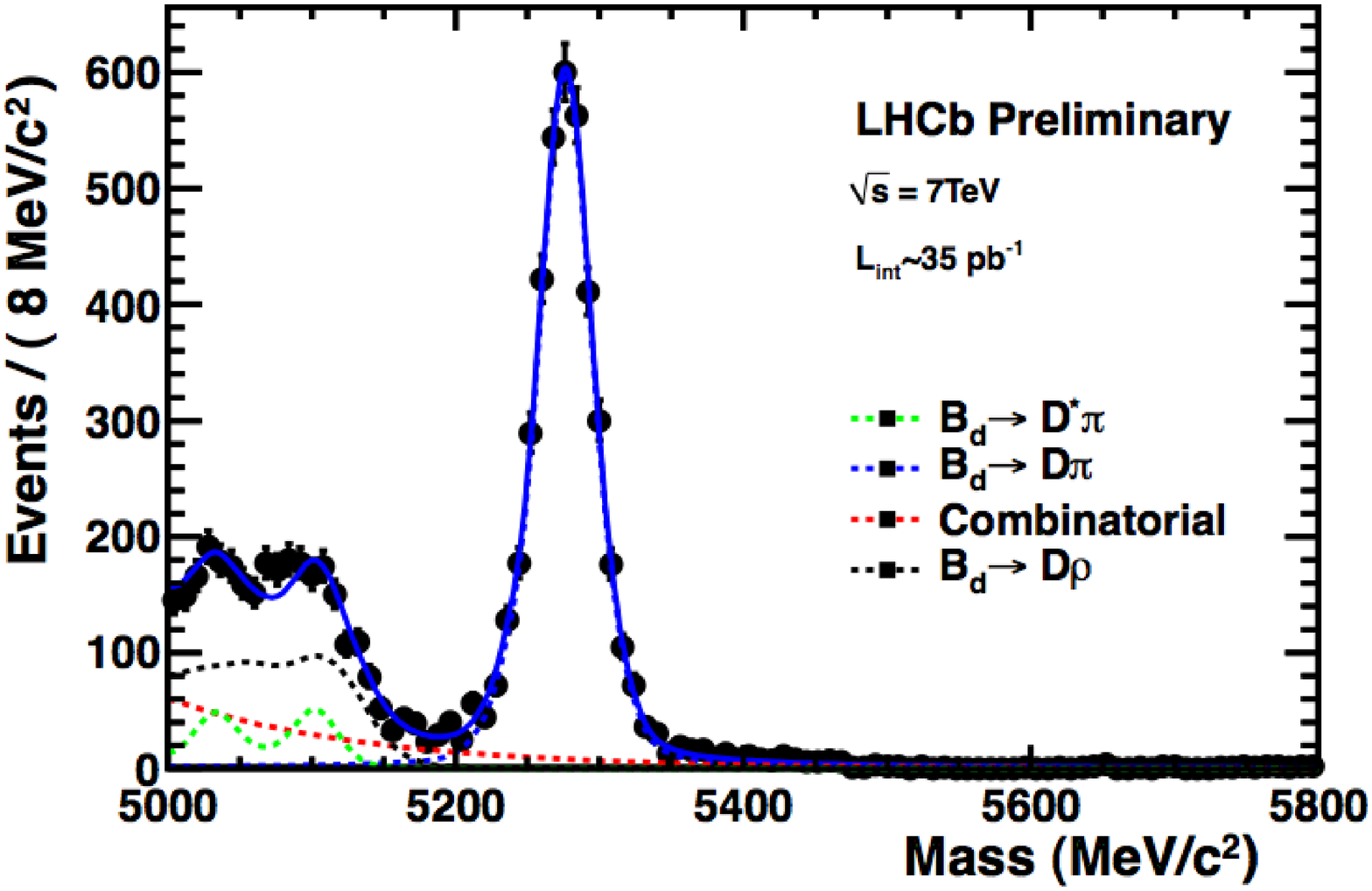}
\caption{Invariant mass distributions for $\btodk$ (left) and $\btodpi$ (right) candidates using 35~$\ipb$ of data.}
\label{fig:b2dh}
\end{figure}

\noindent More details of this analysis can be found in Ref.~\cite{ref:b2dk}.

\section{Measurement of $\br(\bstodzerokstar)$}

The decay $\btodzerokstar$ is of great interest because it can be used in much the same
way as $\btodzerok$ to determine $\gamma$. Although the parent meson is a $B^0$, the final
state is flavor-specific, and therefore the flavor at production is known.
Observing $\btodzerokstar$ will require a larger data set, but a similar decay with larger 
expected rate is $\bstodzerokstar$. This decay, reconstructed in the same final state as 
$\btodzerokstar$, represents a sizeable background, and thus measuring its rate, which
is currently not known, is important.

Its branching fraction is measured relative to the kinematically similar $\btodzerorho$ decay.
The analysis is very similar to the analysis of $\btodk$, except here the bachelor is a vector
resonance. Candidate $K^{*0}\to K^+\pi^-$ ($\rho^0\to\pi^+\pi^-$) decays are required to have 
invariant masses within 50 (150) MeV/$c^2$ of the nominal resonance mass and helicity angle
$\cos(\theta_h)>0.4$. A number of selections are applied, similar to those described above
(see Ref.~\cite{ref:b2d0kstar} for a full list of cuts). Taking advantage of the nearly identical
final states, both L0 TOS and L0 TIS events are used to maximize statistics.
The invariant mass spectra for $\bstodzerokstar$ and $\btodzerorho$
are shown in Fig.~\ref{fig:bstodzerokstar}. Total signal yields of $154\pm14$ $\btodzerorho$ and
$35\pm7$ $\bstodzerokstar$ are observed. This is the first observation of the
$\bstodzerokstar$ decay.

\begin{figure}[ht]
\centering
\includegraphics[width=75mm]{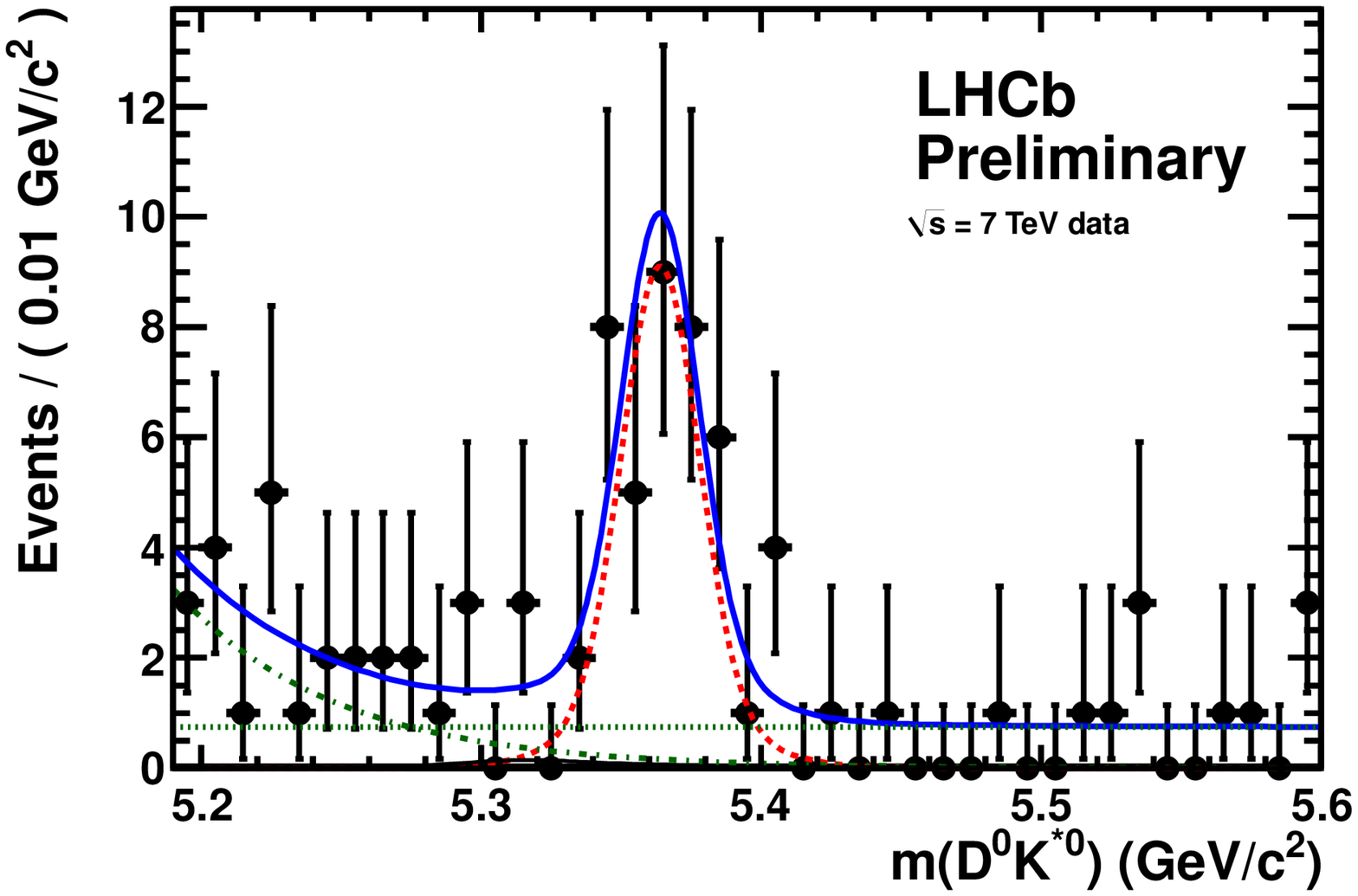}
\includegraphics[width=75mm]{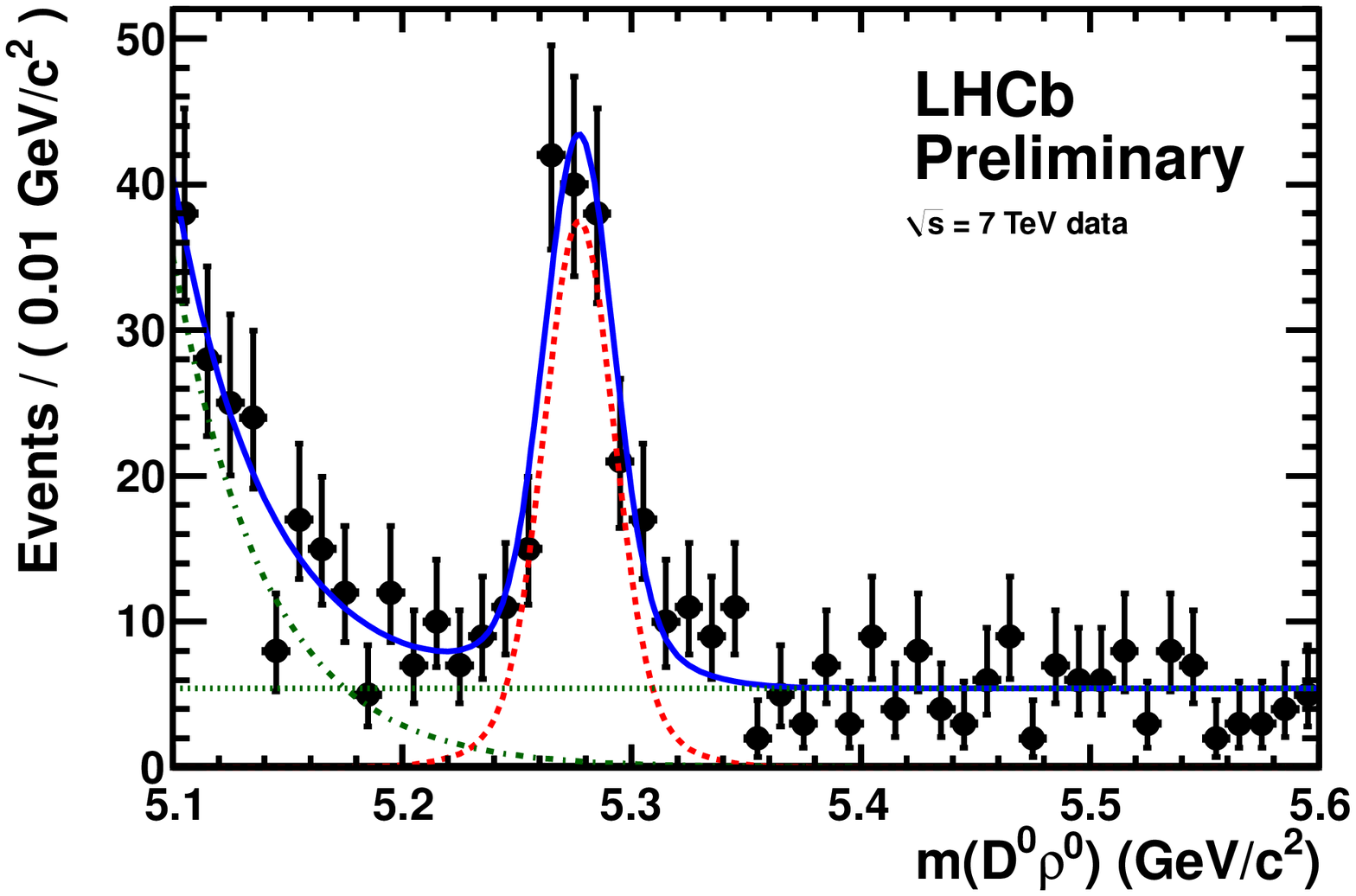}
\caption{Invariant mass distributions for $\bstodzerokstar$ (left) and 
$\btodzerorho$(right) candidates, using 35~$\ipb$ of data.}
\label{fig:bstodzerokstar}
\end{figure}

After correcting by the ratio of efficiencies, the ratio of branching fractions is measured to be:

\begin{equation}
\frac{\br(\bstodzerokstar)}{\br(\btodzerorho)} = 1.39\pm0.31\pm0.17\pm0.18 \nonumber
\end{equation}

\noindent where the uncertainties are statistical, experimental systematic, and the
uncertainty on the $b\to B_d/b\to B_s$ fragmentation fraction.


\section{Measurement of $\br(\btodkpipi)$ and $\br(\btodzerokpipi)$}

In addition to using $\bstodsk$ to measure $\gamma$ in a time-dependent analysis, one can also
make use of $\bstodskpipi$. The branching fraction for the latter is likely to be 2-3 times larger
and will have better proper time resolution. This gain is offset by the lower total efficiency
for observing these decays due to the lower total acceptance, lower average particle $p_T$, etc.
A first step toward observing $\bstodskpipi$ is to observe the CS $\btodkpipi$ and $\btodzerokpipi$ decays.

The selection criteria on the $D^{+,0}$ meson are similar to those described above. The $K^-\pi^+\pi^-$
that accompanies the $D$ meson is reconstructed with similar selections to the $D$ meson, except 
the invariant mass window extends from 0.8-3~GeV/$c^2$. A more detailed description of
the selection criteria are given in Ref.~\cite{ref:cs_decays}. The branching fractions are normalized
to the corresponding CF $\btodpipipi$ and $\btodzeropipipi$ decays~\cite{lathuile_note}. Like $\bstodzerokstar$,
we take advantage of the similar final states and allow for both TOS and TIS events to increase the observed yields.
The invariant mass spectra for $\btodkpipi$ and $\btodzerokpipi$ are shown in Fig.~\ref{fig:btodkpipi} (top plots),
along with the normalization modes (bottom).

\vspace{-0.1in}
\begin{figure}[ht]
\centering
\includegraphics[viewport=0 0 567 540,clip,width=60mm]{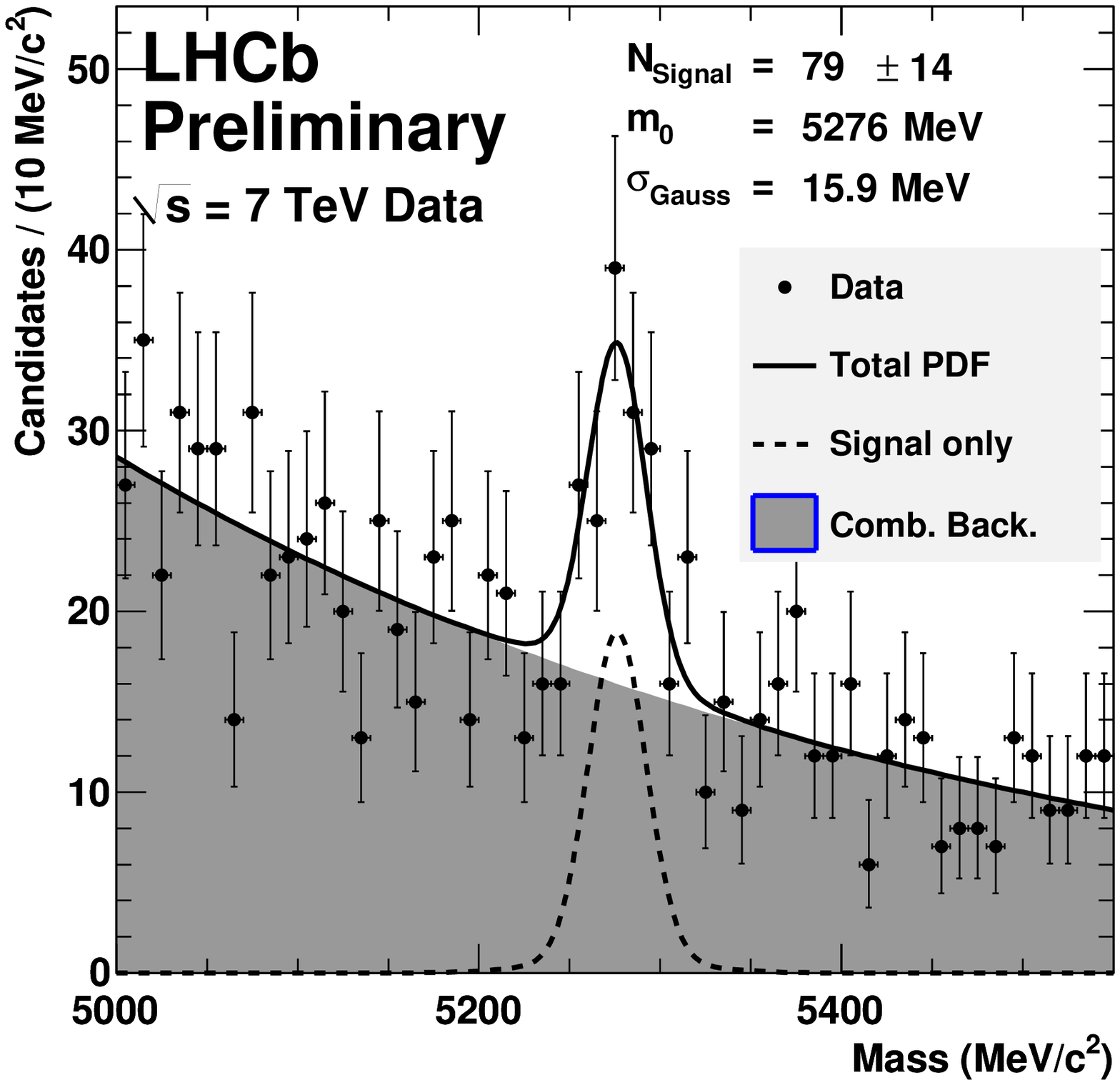}
\includegraphics[viewport=0 0 567 540,clip,width=60mm]{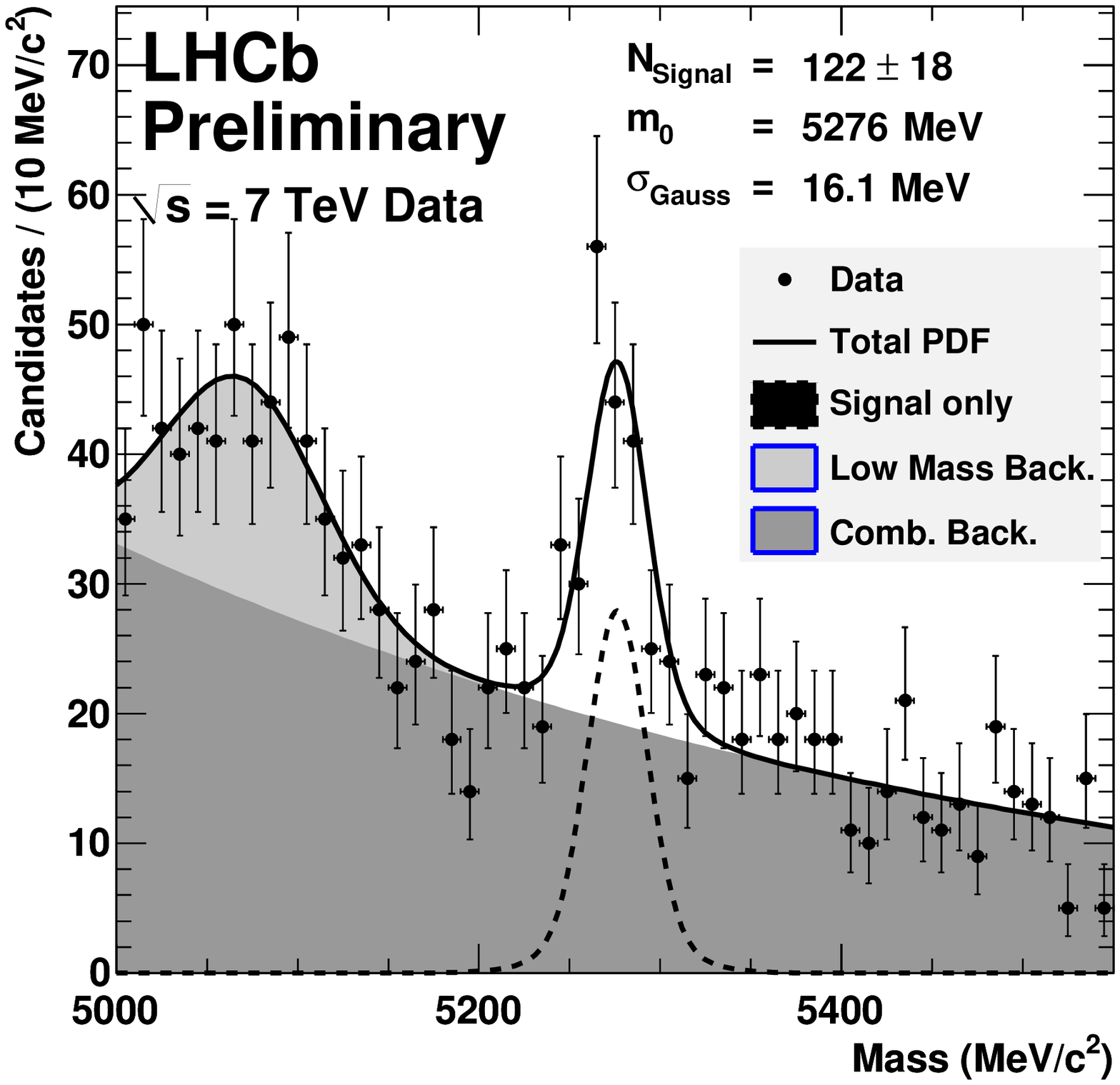}
\includegraphics[viewport=0 0 567 540,clip,width=60mm]{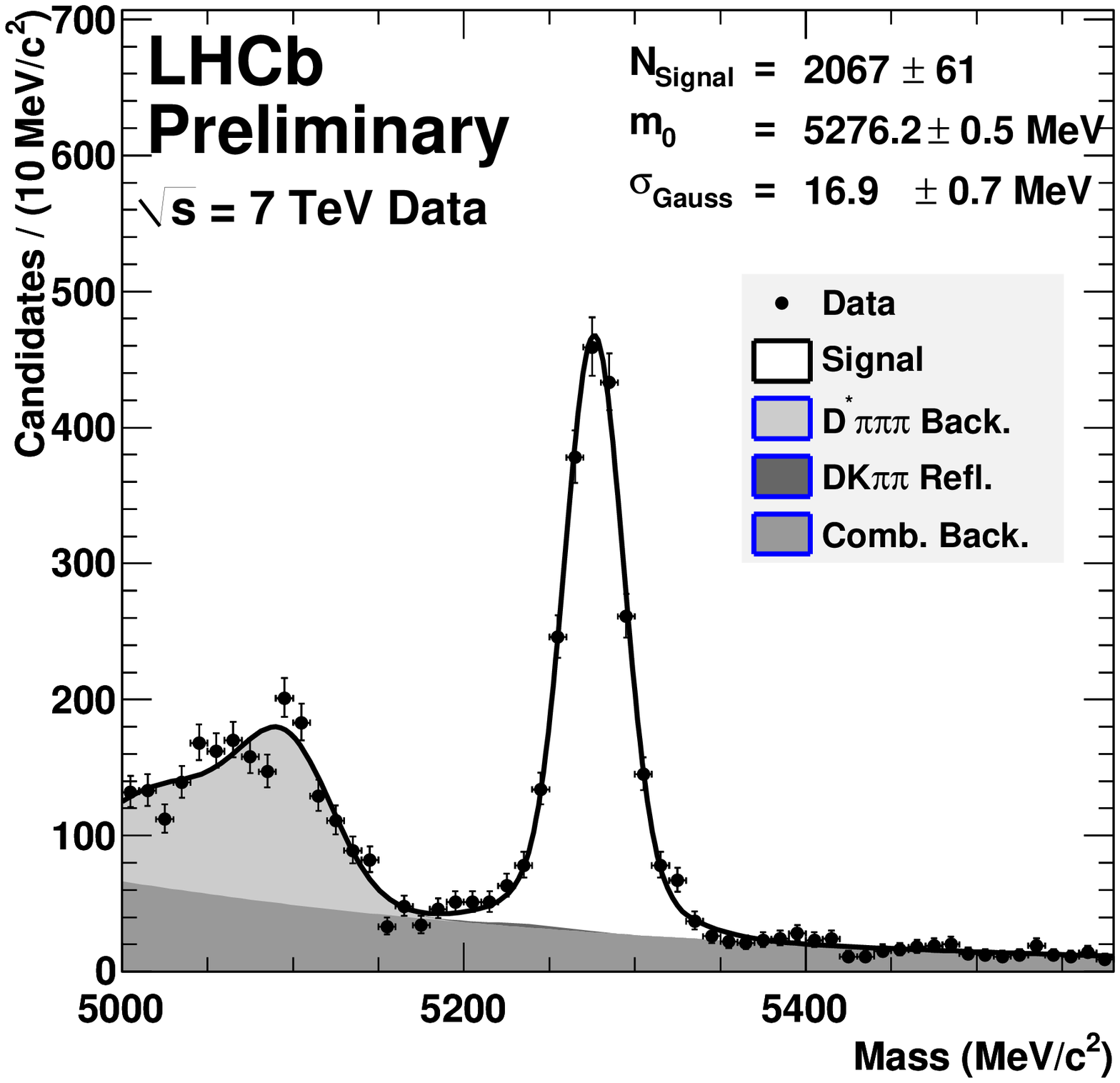}
\includegraphics[viewport=0 0 567 540,clip,width=60mm]{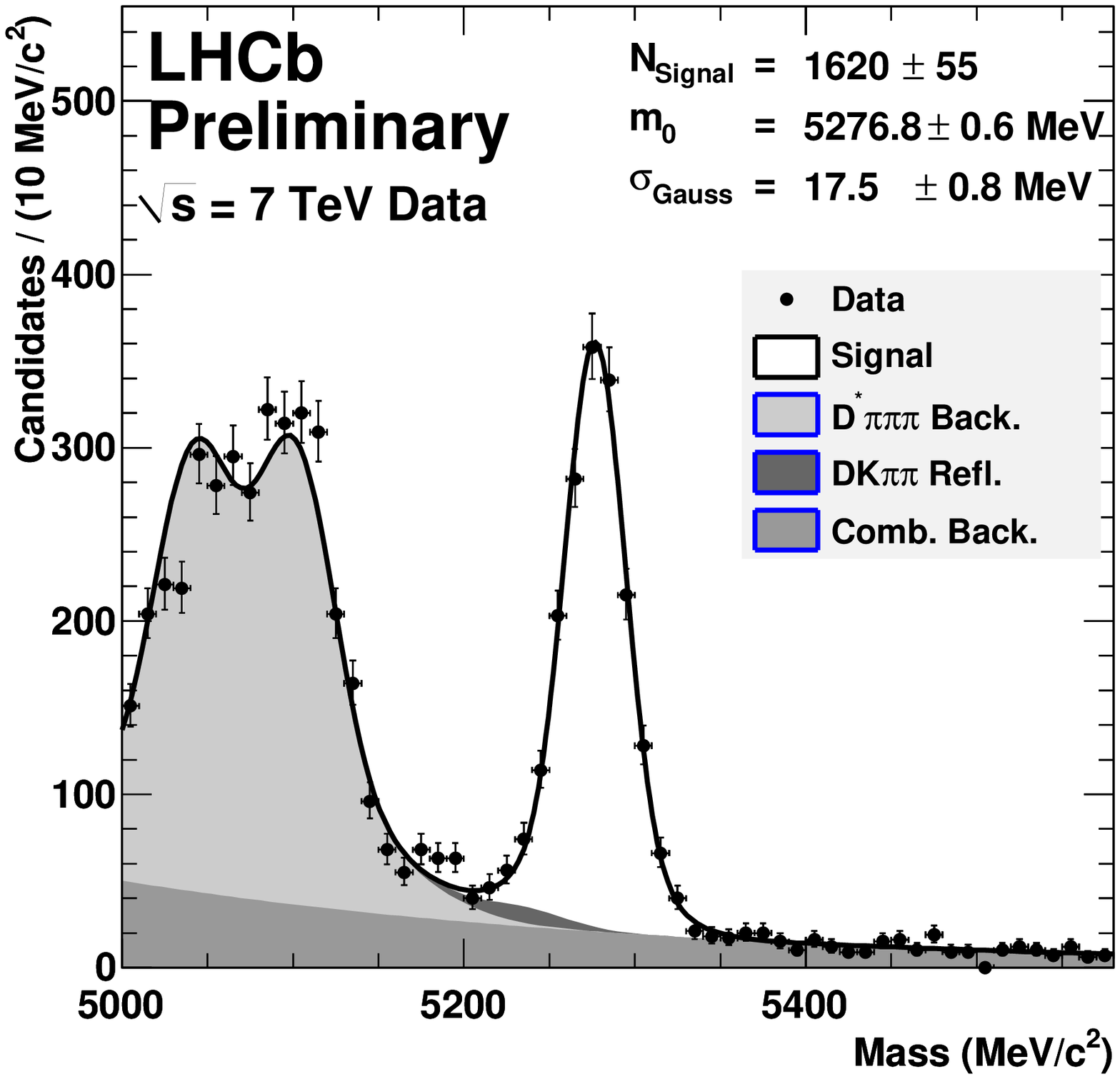}
\caption{Invariant mass distributions for the signal modes, $\btodkpipi$ (top left) and $\btodzerokpipi$(top right),
and for the normalization modes, $\btodpipipi$ (bottom left) and $\btodzerokpipi$(bottom right),
using 35~$\ipb$ of data.}
\label{fig:btodkpipi}
\end{figure}

We observe $79\pm14$ and $122\pm18$ $\btodkpipi$ and $\btodzerokpipi$ decays in the signal modes,
respectively, with corresponding yields of $1620\pm55$ and $2067\pm61$ $\btodpipipi$ and $\btodzeropipipi$ in the
normalization modes. These CS decays are
first observations and have corresponding statistical significances of 6.6 and 8.0 over the background-only hypothesis.
The ratio of efficiencies between the signal and normalization mode are close to unity, as expected. The measured
ratios of branching fractions are found to be:

\begin{eqnarray}
\frac{\br(\btodkpipi)}{\br(\btodpipipi)} &=& (5.2\pm0.9\pm0.5)\% \\
\frac{\br(\btodzerokpipi)}{\br(\btodzeropipipi)} &=& (9.6\pm1.5\pm0.8)\% \nonumber
\end{eqnarray}

\section{Summary}

In summary, we have presented three sets of measurements in $\btodh(h)(h)$ decays made using 35~$\ipb$ of data
from LHCb. Large yields in $B^-\to D^0h^-$ have also been observed (not shown here), which show
great promise for time-independent measurements of $\gamma$.
The observations and measurements of these decays at LHCb provide confidence that we are on track
to carry out the program of $\gamma$ measurements with larger data samples.

\end{document}